\documentclass[twocolumn,aps,prl,superscriptaddress,tightenlines]{revtex4}
\usepackage{amsmath}
\usepackage{amsfonts}
\usepackage{graphicx}
\usepackage{epsfig}
\usepackage{color}
\usepackage{textcomp}
\usepackage[colorlinks,citecolor=blue]{hyperref}

\begin{document}
\title{\emph{N}-Phonon Bundle Emission via the Stokes Process}
\author{Qian Bin}
\affiliation{School of Physics, Huazhong University of Science and Technology, Wuhan, 430074, P. R. China}

\author{Xin-You L\"{u}}\email{xinyoulu@hust.edu.cn}
\affiliation{School of Physics, Huazhong University of Science and Technology, Wuhan, 430074, P. R. China}

\author{Fabrice P. Laussy}
\affiliation{Faculty of Science and Engineering, University of Wolverhampton, Wulfruna Street, Wolverhampton WV1 1LY, United Kingdom}
\affiliation{Russian Quantum Center, Novaya 100, 143025 Skolkovo, Moscow Region, Russia}

\author{Franco Nori}
\affiliation{Theoretical Quantum Physics Laboratory, RIKEN Cluster for Pioneering Research, Wako-shi, Saitama 351-0198, Japan}
\affiliation{Physics Department, The university of Michigan, Ann Arbor, Michgan 48109-1040, USA}

\author{Ying Wu}\email{yingwu2@126.com}
\affiliation{School of Physics, Huazhong University of Science and Technology, Wuhan, 430074, P. R. China}
\date{\today}

\begin{abstract}
  We demonstrate theoretically the bundle emission of $n$ strongly correlated
  phonons in an acoustic cavity QED system. The mechanism relies on
      Stokes resonances that generate super-Rabi oscillations between
  states with a large difference in their number of excitations,
  which, combined with dissipation, transfer coherently pure
  $n$-phonon states outside of the cavity. This process works with
  close to perfect purity over a wide range of parameters and is
  tunable optically with well-resolved operation conditions.  This
  broadens the realm of quantum phononics, with potential applications
  for on-chip quantum information processing, quantum metrology, and
  engineering of new types of quantum devices, such as optically heralded $n$-phonon guns.
\end{abstract}

\maketitle
The manipulation of quantum states is one of
  the main topics of modern science. In the case of photons, an increasingly
  popular research line is that of multi-photon
  physics\,\cite{A. Kubanek,Y. Ota,A. Gonzalez-Tudela,C. S. Munoz1,A. F. Kockum,
    Y. Chang}, with potential applications for multi-photon
  laser\,\cite{D. J. Gauthier}, beating the diffraction
  limit\,\cite{M. DAngelo} and
  metrology\,\cite{I. Afek}. In particular, a scheme
  for the direct generation of $n$-photon states in the same mode
  ($n$-photon bundles) has been recently
  proposed under the platform of cavity
  quantum electrodynamics (cQED)\,\cite{C. S. Munoz1}. 
  
Besides photons, phonons (the quanta of mechanical waves) have emerged
as strong candidates for the engineering of solid-state quantum
devices and on-chip quantum communications, with several distinct
advantages. First, the speed of acoustic waves is significantly slower
than the speed of light, and thus it is more suitable for
communications over short distances, such as a few hundred micrometers
or less (i.e., on-chip communication)\,\cite{M. V. Gustafsson,Mark
  C. Kuzyk}. Second, since phonons can only propagate in a medium,
they are immune to radiation losses into the vacuum. Lastly, phonon
cavities are greatly tunable, with resonant frequency-ranges from
gigahertz (GHz) to terahertz (THz) having already been
fabricated\,\cite{P. Borri,E. M. Weig,G. Rozas,E. Stock,O. O. Soykal,A. Fainstein,Y. T. Xu,P. Kharela}.
THz phonons have wavelengths comparable to the lattice constants,
which have important applications for sensing and nanoscale imaging,
such as detecting microscopic subsurface structures with atomic
precision. Consequently, quantum phononics has progressed enormously,
including the investigation of phonon
lasers\,\cite{H. X. Han,J. Kabuss1,W. Maryam}, phononic quantum
networks\,\cite{M. -A. Lemonde,G. Calajo}, the detection of
electron-phonon interaction in double quantum
dots\,\cite{T. R. Hartke,M. J. Gullans1} and quantum acoustic
devices\,\cite{M. J. A. Schuetz,R. Manenti}. The generation of
multi-phonon quantum states, as a fundamental milestone on the road of
acoustic quantum devices, becomes an important task of phononics. For
example, antibunching bundles and NOON phonon states could be valuable
as $n$-phonon sources\,\cite{Y. W. Chu,K. J. Satzinger} and for
acoustic quantum precision measurements\,\cite{K. Toyoda,J. H. Zhang},
respectively. Multi-phonon processes have also important applications
in ultrasensitive biodetection\,\cite{X.Y. Chu}.

Here, we present a method for implementing $n$-phonon bundle emission
from a quantum dot (QD) coupled to an acoustic nanocavity with
electron-phonon coupling and coherently driven by a laser at the
$n$th-order phonon sideband. This optically-driven Stokes process
realizes super-Rabi oscillations\,\cite{D. V. Strekalov} between
states with large differences in their number of excitations. The pure
bundle-emission can be achieved by opening a dissipative channel for
such super-Rabi oscillations induced by the Stokes
resonances. Compared to the earlier work on $n$-photon bundles
emission~\cite{C. S. Munoz1}, here we have a different physical
mechanism for achieving super-Rabi oscillations, i.e., through the
optically-driven Stokes process. In particular, the QD flip is
accompanied by an $n$-phonon generation in the cavity, induced by the
electron-phonon interaction. In contrast, Ref.\,\cite{C. S. Munoz1}
relies on the excitation of a dressed QD at the $(n+1)$th rung
together with a $n$-quanta energy transfer from the QD to the cavity,
induced by the Jaynes-Cummings interaction, i.e., Purcell-enhancing
the so-called leapfrog transitions. Our work introduces the nonlinear
Stokes process into the theory of bundle emission. This enlarged
regime together with the more complex level structure shows that
$n$-quanta bundle emission is not limited to particular platforms and
configurations but can be exploited in more general settings; in
particular since the ideal Stokes resonance can be realized over a
wide range of parameters.

The different mechanism leads to a series
of exclusive advantages featured by our proposal.  Our implementation
is robust to varying electron-phonon coupling strengths and/or driving
strengths, which only change the resonant conditions. This leaves much
room to achieve $n$-phonon bundle emission and optimize its purity. We
find close to 99\% two-phonon emission and 97\% three-phonon emission
with today's figures of merit\,\cite{G. Rozas, P. Borri, E. M. Weig,
  E. Stock}.  Moreover, here we have the mixed phonon-photon emission,
which allows us to efficiently and conveniently isolate the useful
strongly-correlated $n$-phonon emission from the other (optical)
de-excitation channels. This could be used for the realization of
optical heralded $n$-phonon lasers and guns. Our work opens potential
applications for on-chip quantum communications, e.g., transferring
quantum information with bundles of phonons in future on-chip quantum
networks~\cite{Habraken, Bienfait2019}. It also provides the important
family of nonlinear Stokes processes with a new type of quantum
engineering, besides the widely applied photon and phonon
manipulations, such as the generation of paired
photons\,\cite{V. Balic}, electromagnetically induced
transparency\,\cite{M. D. Eisaman}, sideband cooling of mechanical
oscillators\,\cite{M. Aspelmeyer},
etc.\,\cite{L. Droenner,T. P. Devereaux}.

\begin{figure}
  \centering
  \includegraphics[width=8.5cm]{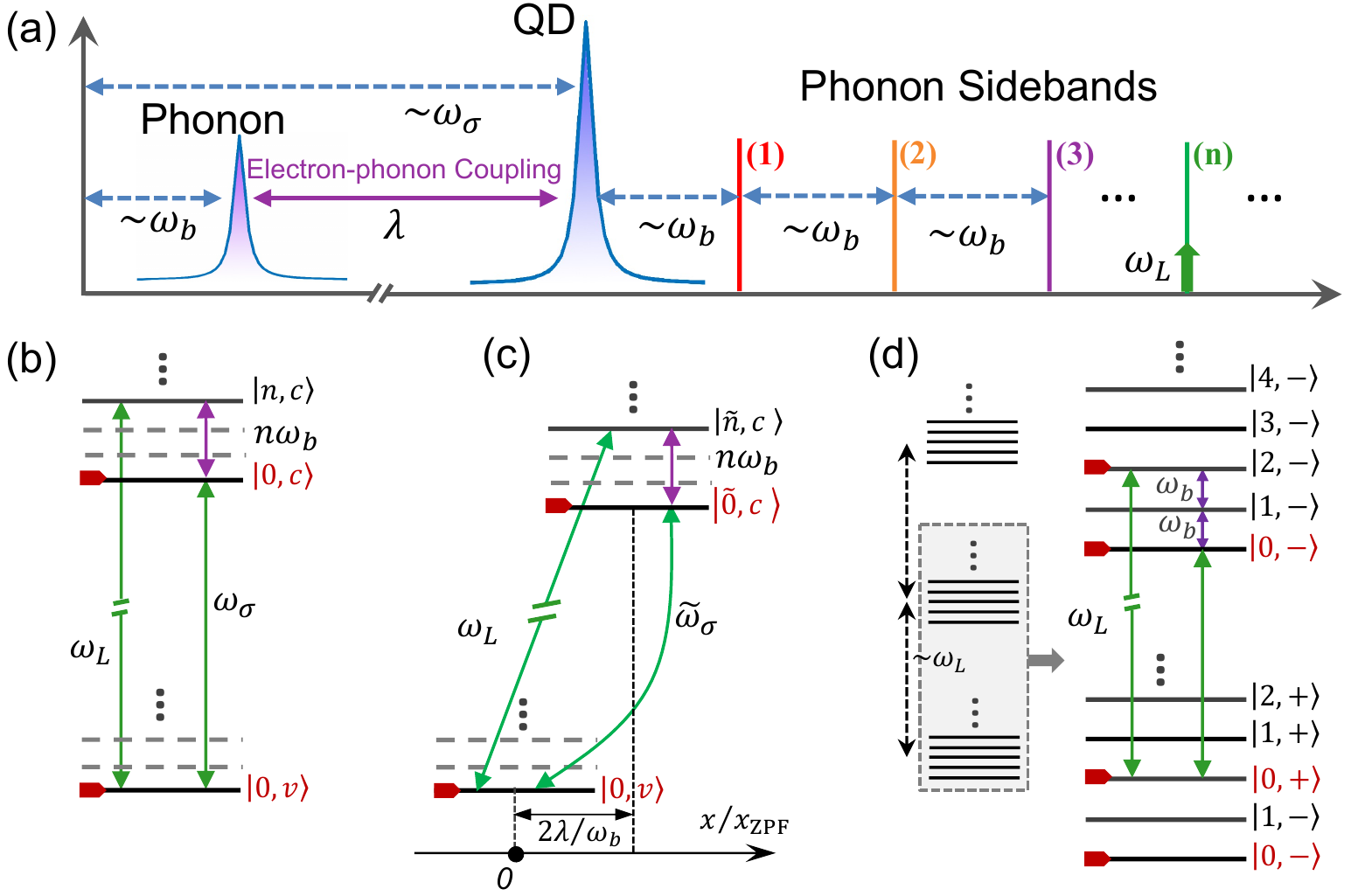}
  \caption{(a) Scheme of the model and of the Stokes resonances
    in the frequency domain. The Stokes resonance is realized when a laser pumps the $n$th-order phonon sideband, which corresponds to (b)
    an ideal Raman process. For large electron-phonon
    coupling and pumping-laser intensity, the energy structure changes
    to that of the strong coupling regime $\lambda\sim\omega_{b}$ (c)
    and of the strong driving (Mollow) regime $\Omega\sim\omega_{b}$
    (d). Panel (d) shows the case of two-phonon resonance, where the states $|n,+\rangle$ and $|n+2,-\rangle$ are degenerate. The dashed arrows in (d) represent the energy gap between two manifolds. Here, $x=(b+b^\dag)/2$ is the position quadrature of the cavity mode, and $x_{\rm ZPF}$ is the corresponding zero-point fluctuation.}\label{fig1}
\end{figure}

\emph{Model and Stokes resonance.}---We consider a phonon cQED model,
with a QD coupled to a single-phonon mode of an acoustic nanocavity
with electron-phonon coupling~$\lambda$, as shown in
Fig.\,\ref{fig1}(a). The QD is a two-level system with conduction-band
state $|c\rangle$, valence-band state $|v\rangle$ and band-gap
frequency $\omega_{\sigma}$. The QD is driven by an optical laser with
frequency $\omega_L$ and amplitude $\Omega$. The system Hamiltonian
reads ($\hbar=1$)
\begin{align}\label{eq1}
  H=&\omega_{b} b^\dag b+\omega_{\sigma}\sigma^\dag \sigma+\lambda \sigma^\dag \sigma (b^\dag+b)\nonumber\\
    &+\Omega\left(e^{i\omega_L t}~\sigma+e^{-i\omega_L t}~\sigma^\dag\right),
\end{align}
where $b^\dag$ ($b$) is the creation (annihilation) operator of the
phonon mode with resonance frequency $\omega_{b}$ and
$\sigma=|v\rangle\langle c|$ ($\sigma^{\dagger})$ is the Pauli annihilation
(creation) operator for the QD. Over a wide range of
    parameters, this system displays Stokes resonances associated
with the periodic generation of $n$~phonons in the acoustic cavity.

\begin{figure}
  \centering
  \includegraphics[width=8.6cm]{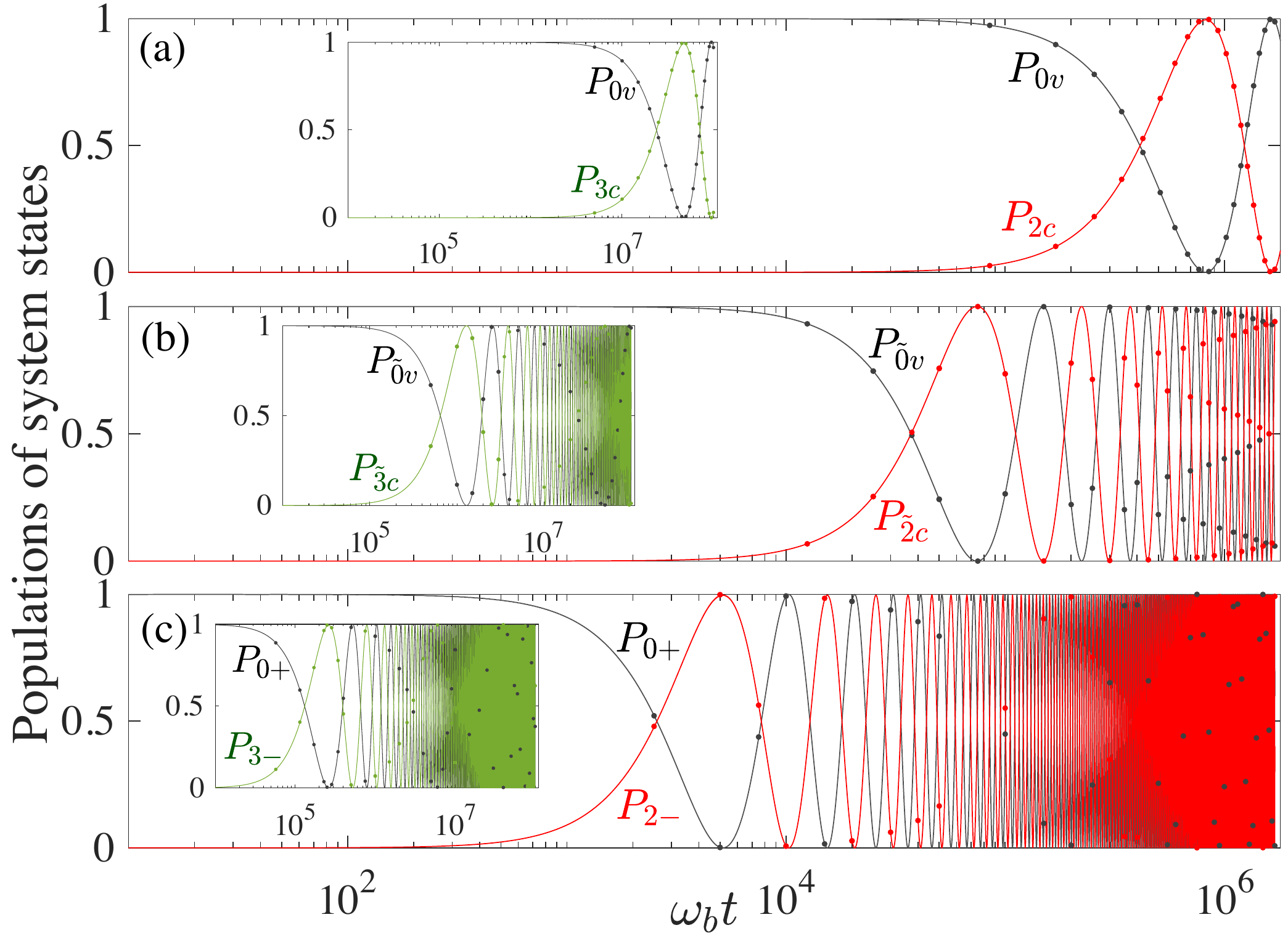}
  \caption{Super-Rabi oscillations as seen through the population dynamics
    $P_{jk}(t)=|\langle j,k|\psi(t)\rangle|^2$ ($j=n,\tilde{n}$ and
    $k=c,v,+,-$) with $n=2,3$ corresponding to the main and insert
    parts, respectively. The solid lines and dots correspond to the
    exact numerical results and the analytical approximate solutions
    based on $\Omega^{(n)}_{\rm eff}$~\cite{Supp}, respectively.
    System parameters are (a) $\lambda/\omega_b=0.03$,
    $\Omega/\omega_b=0.003$, (b) $\lambda/\omega_b=0.1$,
    $\Omega/\omega_b=0.003$, (c) $\lambda/\omega_b=0.03$,
    $\Omega/\omega_b=0.8$, corresponding to the regimes of
    Figs.\,\ref{fig1}(b),\,\ref{fig1}(c) and\,\ref{fig1}(d),
    respectively.
  }\label{fig2}
\end{figure}

First, in the parameter regime $\Omega, \lambda\ll\omega_{b}$, where
the influence on the energy structure of the electron-phonon coupling
and driving laser can be ignored, the eigenstates of the system are
given by the product states~$|n,c/v\rangle$. As shown in
Fig.\,\ref{fig1}(b), the ideal Stokes resonance between states $|0,v\rangle$, $|0,c\rangle$, and
$|n,c\rangle$ is realized when the QD is driven at the frequency of
the $n$th-order phonon sideband, i.e.,
$\Delta=\omega_{\sigma}-\omega_{L}=-n\omega_b$. Specifically, the QD
flip is accompanied by the emission of $n$~phonons into the acoustic
cavity, induced by the electron-phonon interaction. One can obtain by
perturbation theory the approximate Stokes transition rate
between $|0,v\rangle$ and $|n,c\rangle$, leading to their super-Rabi
oscillations at rate $\Omega_{\rm{eff}}^{(n)}=\Omega\left(\lambda/\omega_b\right)^{n}/\sqrt{n!}$~\cite{Supp}.

Second, in the parameter regime $\lambda\sim\omega_{b}$, the strong
electron-phonon coupling changes the energy structure, leading to
different Stokes resonance conditions. This brings the system
Hamiltonian to
\begin{align}
  H=\omega_{b} b^\dag b+\tilde{\omega}_{\sigma}\sigma^\dag \sigma+\Omega[\sigma^\dag e^{-i\omega_L t+\frac{\lambda}{\omega_{b}}(b^\dag-b)}+{\rm H.c.}]
\end{align}
through a displaced transformation $H\rightarrow DHD^\dag$, where
$D=\exp{[(\lambda/\omega_{b})\sigma^\dag \sigma (b^\dag-b)]}$, with
$\tilde{\omega}_{\sigma}=\omega_{\sigma}-\lambda^2/\omega_{b}$ the
rescaled flip frequency of the QD. The reduced Hamiltonian is similar
to that describing a trapped ion\,\cite{Blockley1992}, with $n$-phonon Stokes resonances at
$\Delta=\Delta_{n}(\lambda)=\lambda^2/\omega_{b}-n\omega_b$, as shown
in Fig.\,\ref{fig1}(c), with the product state $|n,c\rangle$ being
replaced by the displaced state
$|\tilde{n},c\rangle=D |n,c\rangle$. The $n$-phonon assisted Stokes transition rate becomes
$\Omega_{\rm{eff}}^{(n)}=\Omega
e^{-\lambda^2/{2\omega_b^2}}(\lambda/\omega_{b})^{n}/\sqrt{n!}$\,\cite{Supp}.

\begin{figure}
  \centering
  \includegraphics[width=8.6cm]{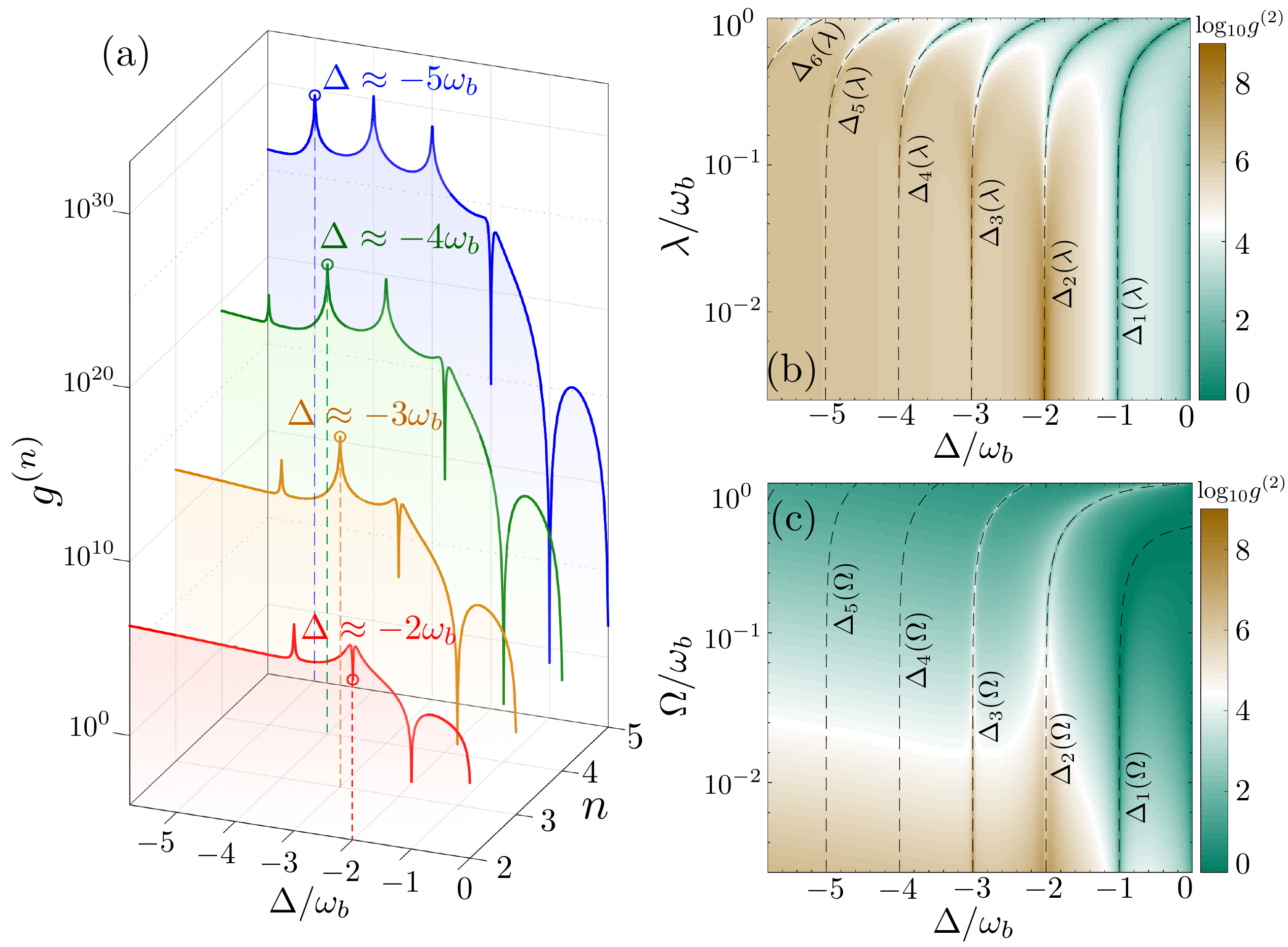}\\
  \caption{(a) Equal-time $n$th-order correlation functions $g^{(n)}$
    as a function of $\Delta/\omega_b$. (b)--(c) Correlation function
    $g^{(2)}$ for different $\lambda/\omega_b$ (b) and
    $\Omega/\omega_b$ (c). The dashed lines indicate the $n$-phonon
    resonances $\Delta=-n\omega_b$, $\Delta=\Delta_{n}(\lambda)$, and
    $\Delta=\Delta_{n}(\Omega)$. System parameters are (a)
    $\lambda/\omega_b=0.03$, $\Omega/\omega_b=0.003$, (b)
    $\Omega/\omega_b=0.003$, (c) $\lambda/\omega_b=0.03$, and
    $\kappa/\omega_b=0.002$, $\gamma/\omega_b=0.0002$ and $\gamma_{\phi}/\omega_b=0.0004$ for panels
    (a)--(c).}\label{fig3}
\end{figure}

Third, in the parameter regime $\Omega\sim\omega_b$, strong driving by
the laser dresses the QD, and forms a Mollow ladder of manifolds,
separated by the energy of the laser. As shown in Fig.\,\ref{fig1}(d),
each manifold consists of many equidistant dressed states
$|n,\pm\rangle$, which is different from the usual Mollow ladder in
the optical cQED
systems\,\cite{B. R. Mollow,C. N. Cohen-Tannoudji,J. Zakrzewski,
  J. C. L. Carreno}. The dressed eigenstates of the QD are
$|\pm\rangle=c_{\pm}|v\rangle\pm c_{\mp}|c\rangle$, where
$c_{\pm}=\sqrt{2}\Omega/(\Delta^2+4\Omega^2\pm\Delta\sqrt{\Delta^2+4\Omega^2})^{1/2}$, with corresponding eigenvalues
$E_{|\pm\rangle}=\Delta/2\pm\sqrt{\Delta^2+4\Omega^2}/2$. In
this regime, $n$-phonon assisted Stokes resonances can still be
realized. This occurs when the laser drives the transition
$|+\rangle\leftrightarrow|-\rangle$, at
$\Delta=\Delta_{n}(\Omega)=-\sqrt{(n\omega_b)^2-4\Omega^2}$. The corresponding $n$-phonon transition rate is $\Omega_{\rm eff}^{(n)}=(-1)^{n}\Omega\left(\lambda/\omega_b\right)^{n}[\prod^{n-1}_{k=1}(nc_-^2-k)]/[(n-1)!\sqrt{n!}]$\,\cite{Supp}.

We illustrate the above discussion in Fig.\,\ref{fig2}, where we
present the essentially perfect super-Rabi oscillations
$|0,v\rangle\leftrightarrow|n,c\rangle$,
$|0,v\rangle\leftrightarrow|\tilde{n},c\rangle$, and
$|0,+\rangle\leftrightarrow|n,-\rangle$, in the absence of
dissipation. This shows that, in these three different regimes, the
two- and three-phonon states are periodically generated with high
fidelity, thanks to the Stokes processes. This is the basic
mechanism for the high-purity $n$-phonon bundle emission. Comparing
Figs.\,\ref{fig2}(a) and \ref{fig2}(b,c), one can see how increasing
$\lambda$ and $\Omega$ speeds up the super-Rabi
oscillations, as is also clear from the approximate analytical
solutions $\Omega^{(n)}_{\rm eff}$ which can be seen to provide an
excellent agreement in all three regimes. As shown below, higher frequencies of oscillations will yield
higher $n$-phonon bundle emission rates.

\begin{figure}
  \centering
  \includegraphics[width=8.6cm]{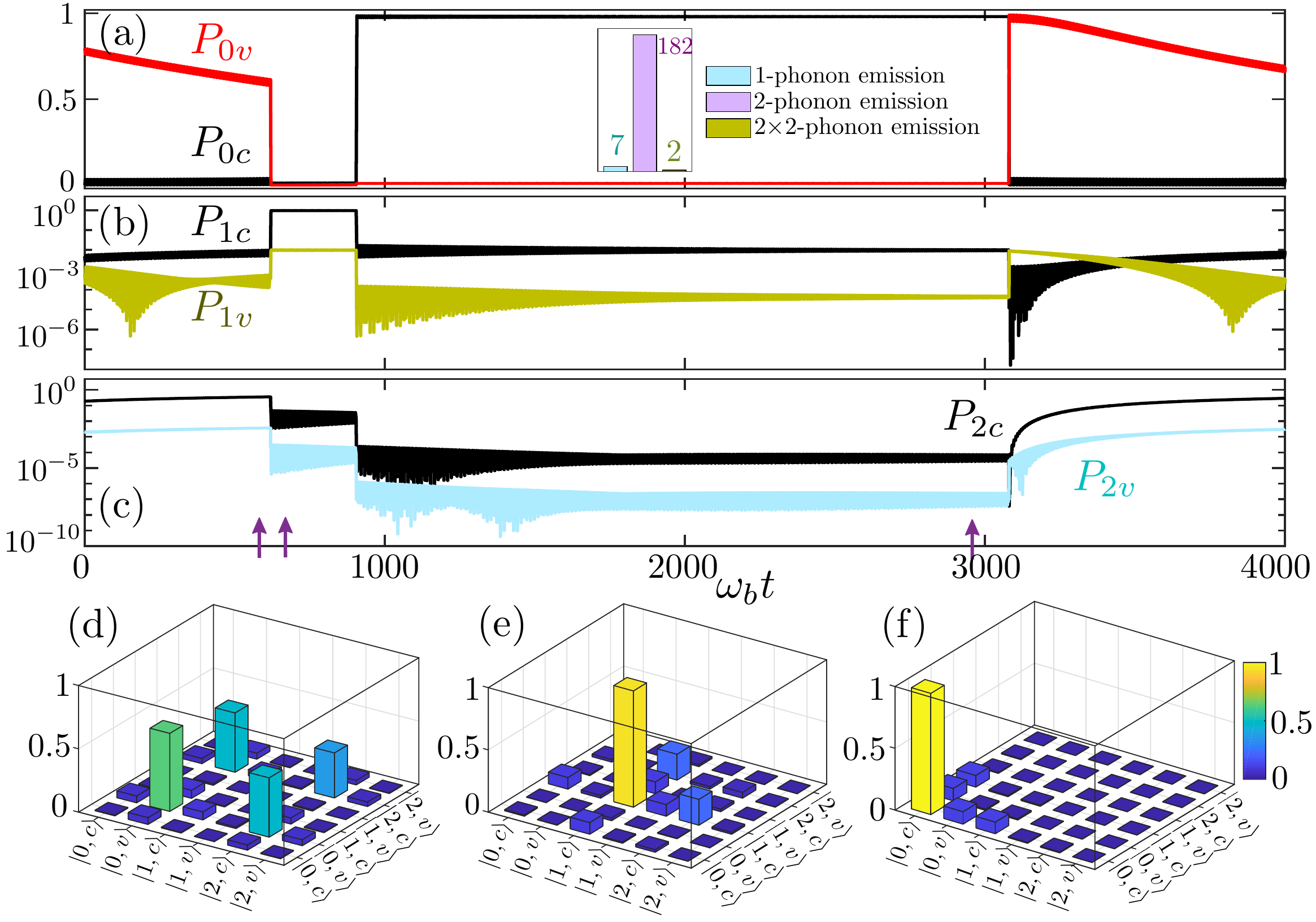}\\
  \caption{(a)--(c) Small fraction of a quantum
    trajectory in the regime of two-phonon
    emission. The insert bar graphs
    show the total number of single-, two- and 2$n$-phonon ($n=2$) clicks over 25
    quantum trajectories. (d)--(f) Full density matrices of the system
    at the times indicated with purple arrows in the trajectory,
    showing the cascade-emission process. System parameters are: $\lambda/\omega_b=0.1$,
    $\Omega/\omega_b=0.2$, $\kappa/\omega_b=0.002$ , $\gamma/\omega_b=0.0002$ and $\gamma_{\phi}/\omega_b=0.0004$, satisfying the regime around sweet point $\kappa\!\approx\!10\Omega^{(2)}_{\rm eff}$\cite{Supp}.}\label{fig4}
\end{figure}

\emph{N-phonon bundle emission.}---The system dissipation has to be
considered to trigger the emission. This is implemented
  with a Lindblad-type master equation
  $d\rho/dt=-i[H,\rho]+\kappa\mathcal{L}[b]+\gamma\mathcal{L}[\sigma]+\gamma_{\phi}\mathcal{L}[\sigma^\dag
  \sigma]$\,\cite{R. J. Glauber}, where
$\mathcal{L}[\emph{O}]=(2\emph{O}\rho\emph{O}^\dag-\rho\emph{O}^\dag\emph{O}-\emph{O}^\dag\emph{O}\rho)/2$,
$\kappa$ ($\gamma$) is the cavity (QD) decay rate, and $\gamma_{\phi}$
is the rate of pure dephasing of the QD.  Dissipation transfers the
above intracavity $n$-phonon states to bundles of strongly correlated
phonons outside of the cavity.  A first unambiguous evidence of strong
correlations of the emitted phonons is given by the equal-time
$n$th-order phonon correlation
$g^{(n)}=\langle b^{\dag n}b^n\rangle/\langle b^\dag b\rangle^n$.
Figure\,\ref{fig3}(a) shows the sharp resonances to all orders of
these correlation functions, clearly associated to the Stokes
resonances $\Delta=-n\omega_b$ in the weak-coupling and low-driving
regime. Note furthermore how a dip inside the bunching
  peak is observed right at the two-phonon resonance rather than a
  superbunching peak, as could be expected for multi-phonon
  emission. The system enters a
  new regime of emission, namely, of strongly correlated bundles.
As shown in Figs.\,\ref{fig3}(b) and \ref{fig3}(c), with increasing
$\lambda$ and $\Omega$, the resonances in $g^{(2)}$ shift along the
curves $\Delta=\Delta_{n}(\lambda)$ and $\Delta=\Delta_{n}(\Omega)$,
respectively. Figure\,\ref{fig3} also shows collectively that the frequency
differences between the $n$ and $(n+1)$-phonon resonances
($\sim\omega_b$) are almost independent of the value of $n$, which is
another clear signature of Stokes resonances (cf. Fig.\,\ref{fig1}).
Interestingly, even for large $n$, the optimum pumping frequency to
realize $n$-phonon emission can be well resolved under the conditions
$\omega_b\gg\kappa,\gamma$, since the off-resonant phonon excitations
are then strongly suppressed. This allows high purity $n$-phonon
emission also for large~$n$\,\cite{Supp}.  Since the order~$n$ of the
bundle can be controlled simply by adjusting the frequency of the
pumping laser, our proposal realizes a versatile optically controlled
multi-phonon source.

The correlation functions $g^{(n)}$ do not guarantee an actual
\emph{n}-phonon emission, although it reveals strong phonon
correlations at the Stokes resonances. This failure is
  manifest from the resonant dip sitting on the superbunching peak
  when the system enters the pure-bundle emission regime, in which
  case $g^{(n)}$ breaks down as single-phonons lose meaning and
  correlations between bundles themselves should be considered instead, which is achieved through generalized
  correlation functions $g_{m}^{(n)}$\,\cite{Supp}. They show that bundles can be
antibunched ($n$-phonon guns), uncorrelated ($n$-phonon laser) or
bunched (thermal states of bundles). To prove \emph{n}-phonon
emission, we turn to Monte Carlo simulations\,\cite{C. S. Munoz1}.  In
this way, one can follow individual trajectories of the system and
record phonon clicks whenever the system undergoes a quantum
jump. Figures\,\ref{fig4}(a-c) show a tiny fraction of a quantum
trajectory during a two-phonon emission (a larger fraction is shown
in\,\cite{Supp}) under the Stokes resonance. In this segment, the
two-phonon state is initially occupied with a probability greater than
$30\%$, while the probability for the one-phonon state is smaller than
$0.1\%$. When the system undergoes a quantum collapse of its
wavefunction triggered by dissipation, it is thus exceedingly more
likely to realize the two-phonon state. This occurs with the emission
of a first phonon that leaves the system in the one-phonon state with
almost unit probability, also highly likely to be emitted during the
cavity lifetime, and thus shortly after the first phonon, completing
the two-phonon bundle emission. As a result, the system has emitted
two strongly-correlated phonons in a very short temporal window.  The
system that is left in the phonon-vacuum and the excited state of the
QD can then resume the cycle after a direct {\it photon} emission
$|0,c\rangle\rightarrow|0,v\rangle$ from the QD flip.  In the next
cycle, the system undergoes the same cascade emission of phonon pair,
each accompanied by a single {\it photon} emission, which can be used
for heralding purposes (e.g., with a delay line).  The emitted single
{\it photon} does not disturb the phonon bundle thanks to their
different nature.  The insert bar graphs in Fig.\,\ref{fig4}(a) show
that the overall two-phonon emission is the largely dominant process
in this regime, chosen as the optimal two-phonon resonant condition
including the influence of the electron-phonon coupling and of the
driving laser. It also shows that the undesired
  2$n$-phonon bundle emission ($n=2$) with a probability close to 1\%
  is negligible~\cite{Supp}.  This corresponds to two-phonon emission rates as
high as $\approx 1.1\times10^9/s$ when $\omega_b/2\pi=1$~THz, for the
chosen values of $\Omega$ and $\lambda$.  The emitted \emph{n}-phonon
bundle has an intrinsic and characteristic temporal structure as a
result of its dynamical character, which corresponds to the
spontaneous emission of a Fock state\,\cite{C. S. Munoz1}. As shown in
Figs.\,\ref{fig4}(d)--\ref{fig4}(f), initially, the system is
essentially in a quantum superposition of the states $|0,v\rangle$ and
$|n,c\rangle$. It then experiences a rapid cascade emission through
the Fock states $|n_{i}\rangle$ ($0\leq n_{i}\leq n$) in a short time
window. Here, $n=2$ and $n_{i}=1$ show the cascaded two-phonon
emission (other cases are shown in\,\cite{Supp}).

\begin{figure}
\centering
\includegraphics[width=8.6cm]{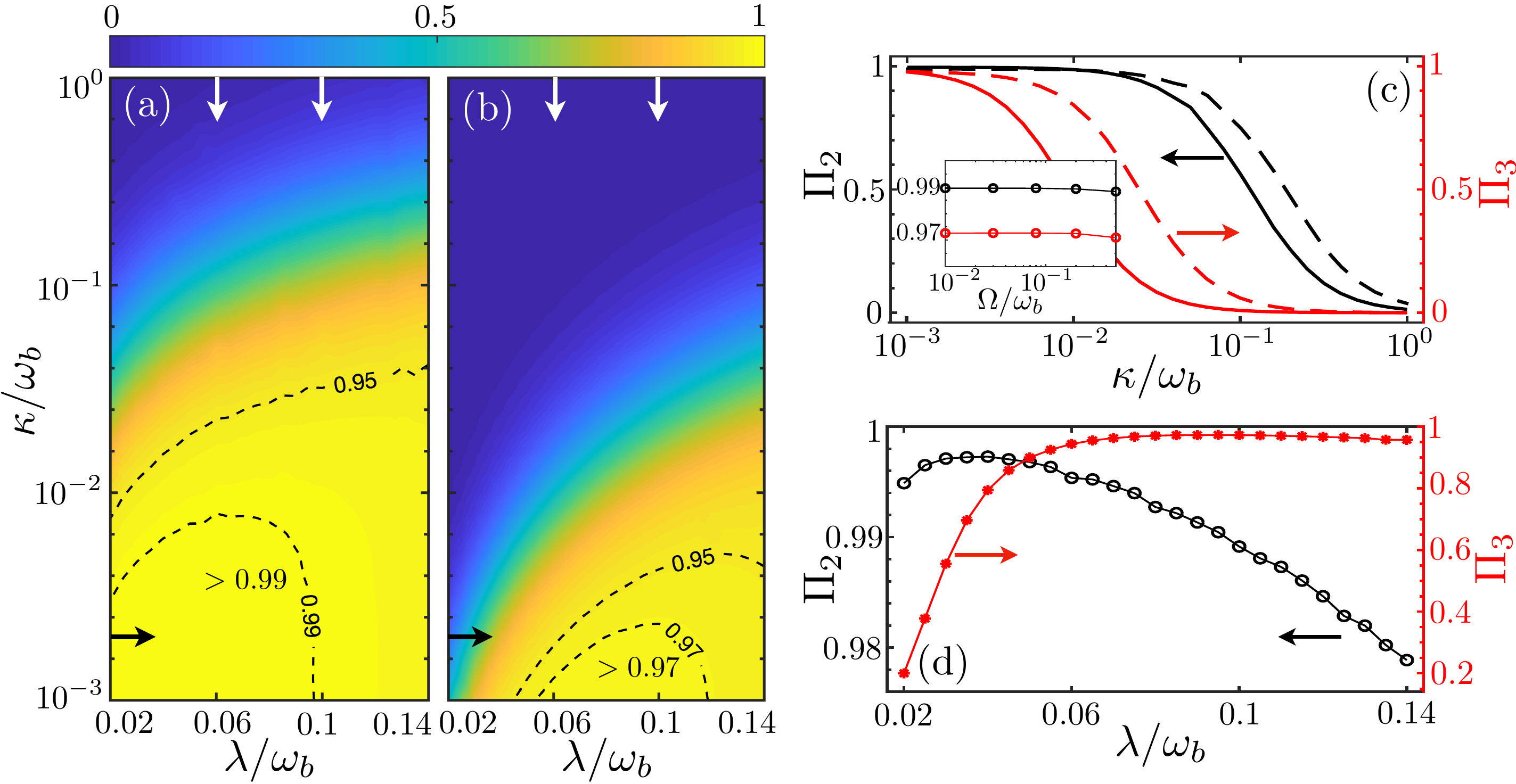}\\
\caption{Purities of (a) two-phonon and (b) three-phonon emissions vs
  $\lambda/\omega_b$ and $\kappa/\omega_b$ when $\Omega/\omega_b=0.2$. (c) Vertical cuts of the purity along
  the arrows in (a) and (b). The solid and dashed lines correspond to
  $\lambda/\omega_b=0.06$ and $\lambda/\omega_b=0.1$,
  respectively. (d) Horizontal cuts of the purity along
  the arrows in (a) and (b).  Inset: Purities vs $\Omega/\omega_b$ when
  $\lambda/\omega_b=0.1$ and $\kappa/\omega_b=0.002$. System parameters are: $\gamma/\omega_b=0.0002$ and $\gamma_{\phi}/\omega_b=0.0004$.}\label{fig5}
\end{figure}

\emph{Discussions of purity and experimental feasibility.}---In an
actual system, or with exact numerical simulations, there is always a
contamination of the $n$-phonon emission by other processes, even when
the system is driven at a perfect $n$-phonon assisted Stokes
resonance. In our case, this comes chiefly from off-resonant
$m$-phonon emission ($m\neq n$). We thus calculate
  numerically the purity $\Pi_n$ of \emph{n}-phonon emission, which is
  defined as $\Pi_n=\bar{P}_{n}/\sum_{n_i=1}^{n}
  \bar{P}_{n_i}$\,\cite{C. S. Munoz1, C. S. Munoz2} where
  $\bar{P}_{n_i}$ is obtained by sampling the phonon-state populations
  in time-windows~$T$ chosen at random times for numerous trajectories
  of the Monte Carlo simulation, until enough statistics is acquired
  to ensure convergence of the probability distribution. Figure\,\ref{fig5} shows that
two- and three-phonon emissions with high purities ($>95\%$) can be
realized in a wide range of parameters for $\kappa/\omega_b$ and
$\lambda/\omega_b$, under the Stokes conditions. The effect of
off-resonant phonon emission on the purity is more notable for larger
cavity decay rates.  Especially, as $n$ increases, this effect is more
obvious due to the occurrence of more off-resonant transitions.
Moreover, increasing $\lambda/\omega_b$ can enhance (decrease) the
three-phonon (two-phonon) emission purity by enhancing the high-order
phonon sideband processes. More interestingly, the inset of
Fig.\,\ref{fig5}(c) shows that high purities ($>97\%$) are robust with
the driving amplitude $\Omega$, since increasing $\Omega$ cannot
extremely enhance the effective Rabi frequency for this
regime. Note that throughout, we have considered
realistic values of pure dephasing for the QD. Its
effect is basically negligible on the purity since
bundles are quickly emitted in a fast cascaded emission following
the wavefunction collapse, prior to which the cavity is in the
vacuum and shielded from QD dephasing~\cite{Supp}.

Regarding experimental implementations, while we have considered here
a semiconductor system with a quantum dot coupled to a THz acoustic
nanocavity, our proposal is not limited to this particular architecture
and could be implemented or adapted in a variety of platforms. For our
specific design, the theoretical model predicts that the current
technology\,\cite{G. Rozas,P. Borri,E. M. Weig,E. Stock} should already
be able to deliver around $99\%$ two-phonon and $97\%$ three-phonon
emission ($\omega_b/2\pi=1\,\rm{THz}$, $\omega_{\sigma}/2\pi=100\,\rm{THz}$,
$\Omega/2\pi=0.2\,\rm{THz}$, $\lambda/2\pi=0.1\,\rm{THz}$, $\kappa/2\pi=2\,\rm{GHz}$, $\gamma/2\pi=0.2\,\rm{GHz}$
and $\gamma_{\phi}/2\pi=0.4\,\rm{GHz}$). A direct observation of our phenomenon
could be made with a nanocalorimeter, where phonons are converted into
electrons by interactions between the nanocavity and the electron gas
(transducer), thus allowing phonon counting and other types of
statistical processing\,\cite{M. L. Roukes,W. D. Oliver} with
well-controlled electronics.

\emph{Conclusions}---We have proposed an efficient method for
producing \emph{n}-phonon bundle emission based on
the Stokes process. The bundle emission with high purity
($>97\%$) is obtained over a wide range of parameters. The purity of
the \emph{n}-phonon emission depends on the cavity decay and the
electron-phonon coupling strength, and is robust with the strength of
the pumping laser. The proposal is easily tunable simply by adjusting
the frequency of the pumping laser, allowing essentially pure two- and
three-phonon emissions with the currently available technology. Our work can also be extended to periodic bundle-emission of phonons by using the collective properties of comb-shaped ensemble of QDs\,\cite{H. S. Dhar}, which is attractive for applications that are clocked in periodic time intervals.

We thank E.~del Valle and C.\,S. Mu\~{n}oz for discussions. We gratefully acknowledge use
of the open source Python numerical packages Numpy, Scipy and
QuTiP\,\cite{J. R. Johansson}. This work is supported by the National Science Foundation of China (Grant Nos.~11822502,
11974125, and 11875029), the National Key Research and Development Program of China grant 2016YFA0301203. F.N. is supported in part by the:
MURI Center for Dynamic Magneto-Optics via the
Air Force Office of Scientific Research (AFOSR) (FA9550-14-1-0040),
Army Research Office (ARO) (Grant No. Grant No. W911NF-18-1-0358),
Asian Office of Aerospace Research and Development (AOARD) (Grant No. FA2386-18-1-4045),
Japan Science and Technology Agency (JST) (via the Q-LEAP program, and the CREST Grant No. JPMJCR1676),
Japan Society for the Promotion of Science (JSPS) (JSPS-RFBR Grant No. 17-52-50023, and JSPS-FWO Grant No. VS.059.18N), the RIKEN-AIST Challenge Research Fund. and the NTT Physics and Informatics Labs.

\end{document}